\documentclass[10pt]{article}
\usepackage{graphicx}
\DeclareSymbolFont{boldletters}{OML}{cmm} {b}{it}
\DeclareSymbolFontAlphabet{\mathbit}{boldletters}
\DeclareMathSymbol{\alpha}{\mathalpha}{letters}{"0B}
\DeclareMathSymbol{\beta}{\mathalpha}{letters}{"0C}
\DeclareMathSymbol{\gamma}{\mathalpha}{letters}{"0D}
\DeclareMathSymbol{\delta}{\mathalpha}{letters}{"0E}
\DeclareMathSymbol{\epsilon}{\mathalpha}{letters}{"0F}
\DeclareMathSymbol{\zeta}{\mathalpha}{letters}{"10}
\DeclareMathSymbol{\eta}{\mathalpha}{letters}{"11}
\DeclareMathSymbol{\theta}{\mathalpha}{letters}{"12}
\DeclareMathSymbol{\iota}{\mathalpha}{letters}{"13}
\DeclareMathSymbol{\kappa}{\mathalpha}{letters}{"14}
\DeclareMathSymbol{\lambda}{\mathalpha}{letters}{"15}
\DeclareMathSymbol{\mu}{\mathalpha}{letters}{"16}
\DeclareMathSymbol{\nu}{\mathalpha}{letters}{"17}
\DeclareMathSymbol{\xi}{\mathalpha}{letters}{"18}
\DeclareMathSymbol{\pi}{\mathalpha}{letters}{"19}
\DeclareMathSymbol{\rho}{\mathalpha}{letters}{"1A}
\DeclareMathSymbol{\sigma}{\mathalpha}{letters}{"1B}
\DeclareMathSymbol{\tau}{\mathalpha}{letters}{"1C}
\DeclareMathSymbol{\upsilon}{\mathalpha}{letters}{"1D}
\DeclareMathSymbol{\phi}{\mathalpha}{letters}{"1E}
\DeclareMathSymbol{\chi}{\mathalpha}{letters}{"1F}
\DeclareMathSymbol{\psi}{\mathalpha}{letters}{"20}
\DeclareMathSymbol{\omega}{\mathalpha}{letters}{"21}
\DeclareMathSymbol{\varepsilon}{\mathalpha}{letters}{"22}
\DeclareMathSymbol{\vartheta}{\mathalpha}{letters}{"23}
\DeclareMathSymbol{\varpi}{\mathalpha}{letters}{"24}
\DeclareMathSymbol{\varrho}{\mathalpha}{letters}{"25}
\DeclareMathSymbol{\varsigma}{\mathalpha}{letters}{"26}
\DeclareMathSymbol{\varphi}{\mathalpha}{letters}{"27}
\DeclareMathSymbol{\Gamma}{\mathalpha}{letters}{"00}
\DeclareMathSymbol{\Delta}{\mathalpha}{letters}{"01}
\DeclareMathSymbol{\Theta}{\mathalpha}{letters}{"02}
\DeclareMathSymbol{\Lambda}{\mathalpha}{letters}{"03}
\DeclareMathSymbol{\Xi}{\mathalpha}{letters}{"04}
\DeclareMathSymbol{\Pi}{\mathalpha}{letters}{"05}
\DeclareMathSymbol{\Sigma}{\mathalpha}{letters}{"06}
\DeclareMathSymbol{\Upsilon}{\mathalpha}{letters}{"07}
\DeclareMathSymbol{\Phi}{\mathalpha}{letters}{"08}
\DeclareMathSymbol{\Psi}{\mathalpha}{letters}{"09}
\DeclareMathSymbol{\Omega}{\mathalpha}{letters}{"0A}

\newcommand{\mbit}[1]{{\mathbit#1}}

\begin{document}    
\setlength{\baselineskip}{18pt}
\title{Critical Couplings in the Nambu--Jona-Lasinio Model
 \\
with the Constant Electromagnetic Fields\thanks{KYUSHU-HET-38} }
\vspace{20ex}

\author{ \\
Masaru~ISHI-I\thanks{isii1scp@mbox.nc.kyushu-u.ac.jp},
Taro~KASHIWA\thanks{taro1scp@mbox.nc.kyushu-u.ac.jp}, and
Naoki~TANIMURA\thanks{tnmr1scp@mbox.nc.kyushu-u.ac.jp}  \\   
\\  Department of Physics, Kyushu University \\ Fukuoka 812-81, 
JAPAN 
}
\maketitle

\vspace{20ex} 

\begin{abstract}
\noindent A detailed analysis is performed for the Nambu--Jona-Lasinio 
model coupled with constant (external) magnetic and/or 
electric fields in two, three, and four dimensions. The infrared cut-off
 is essential for a well-defined functional determinant 
 by means of the proper time method.  
Contrary to the previous observation, the critical coupling remains 
nonzero even in three dimensions. It is also found that 
the asymptotic expansion has an excellent matching with the exact value. 
\end{abstract}

\newpage

The four-fermi interaction model by Nambu and Jona-Lasinio 
(NJL) \cite{NJL} has been discussed to 
investigate the dynamical symmetry breaking (DSB) in a number of cases 
in two, three, and four dimensions. Especially interesting situations are 
found such that NJL is coupled to external sources, which enables us 
to peep into detailed structures of DSB, giving informations of
the chiral symmetry breaking in the QCD vacuum, the planar 
($2+1$-dimensional) dynamics in solid state physics, or the early 
universe when coupled to a curved space-time \cite{IMO}.  
The NJL model minimally coupled to the electromagnetic fields is 
discussed  yielding the result that the electric field 
destabilizes DSB but the magnetic field stabilizes it \cite{KL}. 
In the pure magnetic 
field case, Gusynin, Miransky, and Shovkovy made detailed discussions to
find that there occurs the mass generation even at the weakest 
attractive interaction \cite{GMS1} (in $2+1$ dimensions) and emphasize it 
by means of the dimensional reduction 
\cite{GMS2}. This implies that {\em the critical coupling is zero even if 
the applying magnetic field is infinitesimal} (in the $2+1$-dimensional 
case) which might however 
contradict a na\"{\i}ve expectation. 
The motivation for this work lies here.

We start with the partition function of the NJL model minimally 
coupled to the external electromagnetic field in $D$ (space-time) 
dimensions:
\begin{eqnarray}
	 Z[A] \!\! &\equiv & \!\! 
	 \int {\mathcal D} \psi {\mathcal D} \overline\psi \exp \left[ 
- \int d^{D}x  \Bigl[ \overline\psi \Bigl\{ \gamma_{\mu} \bigl(
\partial_{\mu}
	- iA_{\mu}\bigr) \Bigr\} \psi  - { g^{2} \over 2} \left\{ 
	\bigl( \overline\psi 
	\psi \bigr)^{2} + \bigl(\overline\psi i\gamma_{5}  \psi \bigr)^{2} 
	 \right\} \Bigr] \right] 	\nonumber  \\ 
  \noalign{\vskip -1ex}	
  \!\! & \!\! & \!\!      \\      
   \noalign{\vskip -1ex}  
    \!\! & =  & \!\!  \int {\mathcal D} \sigma {\mathcal D} \pi  {\mathcal D} \psi {\mathcal D}
       \overline\psi \exp \left[ - \int d^{D}x  \Bigl[ {1 \over 
       2g^{2}} \bigl( \sigma^{2} + \pi^{2} \bigr) + 
       \overline\psi \Bigl\{ \gamma_{\mu} \bigl( \partial_{\mu}
	- iA_{\mu}\bigr)  + \bigl(  \sigma + i \pi \gamma_{5}\bigr) 
	\Bigr\}  \psi \Bigr] \right] 
	\  ,  \nonumber 
\end{eqnarray}
where the electromagnetic coupling constant has been absorbed in the 
definition of $A_{\mu}$ and the auxiliary fields, $\sigma$ and $\pi$, have 
been introduced as usual.  The Euclidean metric has been employed.
The fermionic integration gives the 
functional determinant: (this can be considered as the definition 
of the fermionic functional 
measure if some calculative way would be 
provided as in the follows:)
\begin{equation}
 \int {\mathcal D} \psi {\mathcal D}
       \overline\psi \exp\left[ \overline\psi \Bigl\{ \gamma_{\mu} 
       \bigl( \partial_{\mu}
	- iA_{\mu}\bigr)  + \bigl(  \sigma + i \pi \gamma_{5}\bigr) 
	\Bigr\}  \psi  \right] \equiv 	\det \Bigl[ 
	\gamma_{\mu} \bigl( \partial_{\mu}
	- iA_{\mu}\bigr)  + \bigl(  \sigma + i \pi \gamma_{5}\bigr) 
	\Bigr]   \ .
\end{equation}
We then perform the semiclassical approximation, 
that is, shift $\sigma \rightarrow m + \sigma' ; \  \pi \rightarrow \pi'$ 
and assign 
$\sigma'$ and  $\pi'$ as the new integration variables to find  
\begin{equation}
	Z[A]=
	\exp \left[ - VT { m^{2}\over 2g^{2}} + {\rm Tr} \ln \Bigl[ 
	\gamma_{\mu} \bigl( \partial_{\mu}
	- iA_{\mu}\bigr)  + m \Bigr] \right]   
	\Bigl( 1 + O(\mbox{$2$-loop}) \Bigr)  	
\equiv{\mathrm e}^{-VT {\mbox{\Large$v$}}_{\rm T}}  
          \Bigl( 1 + O(\mbox{$2$-loop}) \Bigr)   \ ,                                
\label{totpot}
\end{equation}
where $V$ is the $(D-1)$-dimensional volume of the system 
and $T$ is the Euclidean time 
interval and the trace operation, designated by Tr,
must be taken with respect to the 
space-time as well as the gamma matrices, meanwhile tr implies the 
trace for the gamma matrices. We call ${\mbox{\Large$v$}}_{\rm T}$ the total 
potential. Finally it should be understood that the terms
 of $O(\mbox{two-loop})$ 
are composed by the $\pi' , \sigma'$ integrations.

The functional determinant, 
$I(D)\equiv{\rm Tr}\ln\Bigl[\gamma_{\mu}\bigl(\partial_{\mu}-iA_{\mu}\bigr)+m\Bigr]$ 
can be defined, as is mentioned above, with the aid of 
the proper time method, and is exactly calculable when
the field strength is constant \cite{IZ}:
\begin{equation}
	I(D) =  -  { VT\over 2} 
	\lim_{s \rightarrow 0} { 1 \over (4 \pi)^{D/2}} 
	\int_{0}^{\infty} d\tau 
	\  \tau^{s-D/2 -1} e^{-\tau m^{2}} \left[ \det 
	\Bigl( {\sin \tau F \over \tau F} \Bigr)  \right]^{-1/2} {\rm tr}
	 \   \exp 
	\left( {\tau \over 2} \sigma_{\mu \nu} F_{\mu \nu} \right) 
	\ ,
	 \label{I}
\end{equation}
where $F$ stands for $D \times D$ matrix $(F_{\mu \nu})$ and
 $\sigma_{\mu \nu} = 
\bigl[ \gamma_{\mu}, \gamma_{\nu} \bigr]/2i$. It should be noted that 
(\ref{I}) holds even for odd dimensions, $D=3$: there is no chiral 
transformation but by introducing an additional fermion
 we have a four-component
 theory to be able to discuss the chiral symmetry in a 
parallel manner as in $D=4$ or $2$ \cite{ABKW}. Calculating the 
determinant and the trace, we obtain
\begin{equation}
	I(D) = -VT { {\rm C}^{(D)} F_{D} \over \pi^{D/2} }
	\lim_{s \rightarrow 0}   \int_{0}^{\infty} d 
	\tau  \   \tau^{s -D/2} e^{-\tau m^{2}} \coth \tau F_{D}  
	\label{det}
\end{equation}  
with
\begin{equation}
  F_{D}\equiv\left\{   
    \begin{array}{cc}
      {\displaystyle E\ ,}&{\displaystyle D=2\ ,}\\ 
         \noalign{\vspace{1ex}}
      {\displaystyle\sqrt{B^{2}+\mbit{E}^{2}}\ ,}&  
      {\displaystyle D=3\ ,}\\
         \noalign{\vspace{1ex}}
      {\displaystyle\sqrt{\mbit{B}^{2}+\mbit{E}^{2}}\ ,} & 
         {\displaystyle D=4\ ,}
    \end{array}\right. 
\hspace{20pt}{\rm C}^{(D)}\equiv\left\{   
    \begin{array}{cc}
        {\displaystyle{1\over 4}\ ,}&{\displaystyle\ D=2,\ 3\ ,}\\
           \noalign{\vspace{1ex}}
         {\displaystyle{1 \over 8}\ ,}&{\displaystyle D=4 \ .}  
    \end{array}    \right. 
\end{equation}
However it should be noted that in the four-dimensional case we have 
assumed that $\mbit{B} \cdot \mbit{E} =0$ 
otherwise we have
\begin{equation}
	I_{\rm exact}(D=4) =  
	- VT \lim_{s \rightarrow 0} { F_{+}F_{-} \over 8 \pi^{2} }
	\int_{0}^{\infty} d 
	\tau  \  \tau^{s -1} 
	e^{-\tau m^{2}} \coth \tau F_{+}\coth \tau F_{-} \ ,
	 \label{fullpot}
\end{equation}
where $
	F_{\pm} \equiv
	\left\{ \bigl|  \mbit{B} + \mbit{E} \bigr| \pm 
	\bigl|  \mbit{B} - \mbit{E} \bigr| \right\}/2$.
(A detailed calculation will be published elsewhere \cite{IKT}.)

In this paper we will confine ourselves in the case $\mbit{B} \cdot \mbit{E} =0$, 
therefore, in (\ref{det}) for brevity. Although the integral has entirely 
been regularized if an analytic continuation is made for $s$, 
it is better to introduce an ultraviolet
cut-off $\Lambda$ 
with dimension in order to grasp a physical situation as is done 
in the ordinary gap equation \cite{NJL}. 
Moreover an {\em infrared cut-off must be necessary} in this case 
because we know that there arises an infrared 
divergence when external fields are coupled to the massless 
state.  A more careful treatment is therefore required for the discussion 
on the transition from the massless to the massive state under 
external fields. We then consider instead of (\ref{det}) 
\begin{equation}
	I_{\rm r}(D) \equiv -VT { {\rm C}^{(D)} F_{D} \over \pi^{D/2} } 
	\int_{ 1 \over \Lambda^{2}}^{\infty} d 
	\tau  \   \tau^{-D/2} e^{-\tau ( m^{2} + \epsilon )} 
	\coth \tau F_{D} \ , 
	\label{regdet}
\end{equation}
where $\Lambda$ is the ultraviolet cut-off and 
$\epsilon$ is the infrared one to ensure the existence of the massless 
limit: $m^{2} \longrightarrow 0$. It should be noted that they are 
gauge and Lorentz invariant. With these regularizations
 the integral 
(\ref{regdet}) now becomes well-defined at any time. 
Now the total potential in (\ref{totpot}) reads
\begin{equation}
	{\mbox{\Large$v$}}_{\rm T} = {m^{2} \over 2g^{2}} + 
	{ {\rm C}^{(D)} F_{D} \over \pi^{D/2} } 
	\int_{ 1 \over \Lambda^{2}}^{\infty} d 
	\tau  \   \tau^{-D/2} e^{-\tau ( m^{2} + \epsilon )} 
	\coth \tau F_{D} \ , 
\end{equation}
whose explicit value can be estimated such that
\begin{eqnarray}
 {\mbox{\Large$v$}}_{\rm T} &  = & {m^{2} \over 2g^{2}} +  {m^{2} + 
\epsilon  \over 
4 \pi } \left[  \gamma + 
\ln { m^{2} + \epsilon \over \Lambda^{2} }
 - \ln a + { 1 \over 2a} \bigl( \ln { a 
\over 2 \pi } + 2 \ln \Gamma(a) \bigr) \right] \ ;   
  \hspace{10.5ex} D=2 \ , 
	 \label{pottwo}
	  \\
	  \noalign{\vspace{2ex}}
{\mbox{\Large$v$}}_{\rm T}   &  = &{m^{2} \over 2g^{2}} -  { \left( m^{2} + 
\epsilon \right)  \Lambda \over  2 \pi^{3/2} } -     { \left( m^{2} + 
\epsilon \right)^{3/2}   \over 
2 \pi^{3/2} } \left[   { \sqrt{ \pi } \over 2 a } - { \sqrt{ \pi } \over  
a^{3/2} } \zeta\! \left( - {1 \over 2} , a \right) \right] \ ;  
   \hspace{14.5ex} D=3 \ , 
	\label{potthree} 
	 \\
	 \noalign{\vspace{2ex}}
 {\mbox{\Large$v$}}_{\rm T} &  =  & {m^{2} \over 2g^{2}} - 
 {\bigl( m^{2} + \epsilon  \bigr) \Lambda^{2} \over 8 \pi^{2}}  
+ {\bigl( m^{2} + \epsilon  \bigr)^{2}\over 16\pi^{2}}  
 \left[      \left(  1 + { 1 \over 6a^{2} } \right)
\left(  1 - \gamma - \ln { m^{2} + \epsilon \over \Lambda^{2} } + \ln a 
\right) \right. 
	\nonumber
	 \\
      &  & 
       \label{potfour} 
	 \\
    &   &  \hspace{48ex}  \left. - { 1 \over a } \ln a -{ 2\over a^{2} } \zeta'
            \!\! \left( - 1 , a \right) \right] \ ;  \ \ 
             D=4 \ ,   \nonumber  
 \end{eqnarray}
where $a \equiv(m^{2} + \epsilon)/2F_{D}$, 
$\Gamma(a)$, $\gamma$ is the gamma function, Euler's 
constant respectively, $\zeta(s, a)$ is Riemann's zeta 
function, and $\zeta'(s, a) \equiv d \zeta(s, a) / ds$. We have 
discarded terms of $O\bigl(m^{2}/ \Lambda^{2}\bigr), 
O\bigl(F_{D}/ \Lambda^{2}\bigr),$ and also purely $\Lambda$-dependent 
terms to arrive at the expressions; (\ref{pottwo}) $\sim$ (\ref{potfour}). 

Since our interest is to know the change of the system
undergone by switching on the external field, the parameter, $a$, can be 
considered very large, $a \sim \infty$. Therefore the asymptotic 
expansion for $\ln \Gamma(a), \ \zeta(s, a)$ can be utilized to give
\begin{eqnarray}
	 {\mbox{\Large$v$}}_{\rm T} &  
	 \stackrel{a\rightarrow \infty}{\longrightarrow } & { m^{2} \over 2g^{2}}  +
	 {  m^{2} +  \epsilon  \over 4 \pi }  \left( \gamma + 
	 \ln {   m^{2} +  \epsilon \over \Lambda^{2} } -1 \right)  +  	
	{ F_{2}^{2} \over 
12 \pi \left( m^{2} + \epsilon  \right) } \  ;  \hspace{29ex} D=2 \ , 
\label{asptwo}  \\
{\mbox{\Large$v$}}_{\rm T} & \stackrel{a\rightarrow \infty}{\longrightarrow }
& {m^{2} \over 2g^{2}}- { \left(  m^{2} + 
\epsilon \right) \Lambda  \over 2 \pi^{3/2} } + { \left(  m^{2} + 
\epsilon \right)^{3/2} \over 3 \pi } + { F_{3}^{2} \over 12 \pi \left(  m^{2} + 
\epsilon \right)^{1/2}  } \  ;  \hspace{28.5ex}   D=3 \ ,	
	\label{aspthree}  \\
{\mbox{\Large$v$}}_{\rm T} & \stackrel{a\rightarrow \infty}{\longrightarrow }
& {m^{2} \over 2g^{2}} - { \left(  m^{2} + 
\epsilon \right) \Lambda^{2}  \over 8\pi^{2} } + { \left(  m^{2} + 
\epsilon \right)^{2} \over 16 \pi^{2} } \left( {3 \over 2} -  \gamma
- \ln { m^{2} + \epsilon \over \Lambda^{2} } \right) - { F_{4}^{2} \over 
24 \pi^{2} } \left( \gamma +  \ln { m^{2} + \epsilon \over \Lambda^{2} }
\right)  \  ;    D=4 \ .	
	\label{aspfour}
\end{eqnarray} 
It should be stressed that owing to the infrared cut-off, $\epsilon$, 
we can rely on the asymptotic expansion even in the case $m=0$: here we 
regard the infinitesimal size of external fields as, $ F_{D} << 
\epsilon $.

The gap equation is therefore obtained for $m^{*} \neq 0$ as
\begin{eqnarray}
\hspace{10ex}
	- { 2 \pi  \over g^{2} }  &  =   &  \gamma +\ln x  - { { \mathcal F}_{2}^{2} 
	\over 3 x^{2} } \  ;    \hspace{30ex} D=2  \  , 
	\label{gaptwo} 
	 \\
\hspace{10ex} 
1 -{ \pi^{ 3/2 } \over g^{2} \Lambda }  & = &  \pi^{1/2} \left[ 
x^{1/2} - { { \mathcal F}_{3}^{2} \over 12 x^{3/2} } \right] \ ;  
\hspace{23ex}  D=3 \ , 
	\label{gapthree}
	  \\
\hspace{10ex} 
 1 - { 4\pi^{ 2 } \over g^{2} \Lambda^{2} }	 & = &  x \left( 1 - 
\gamma - \ln x \right) + { { \mathcal F}_{4}^{2}  \over 3 x} \ ;  
\hspace{23ex} 
   D=4 \  ,
	\label{gapfour} 
	\end{eqnarray}
where we have introduced the dimensionless quantities; 
$x\equiv({m^{*}}^{2} + \epsilon)/\Lambda^{2}$, 
${\mathcal F}_{D}\equiv F_{D}/\Lambda^{2}$. 
(Recall that the mass-dimension of the gauge field is always one 
because of the inclusion of the coupling constant.)
Moreover the stability condition, 
$\partial^{2}{\mbox{\Large$v$}}_{\rm T}/\partial m^{2}\Bigr|_{m^{*}}\geq 0$, 
must be fulfilled; which reads $df(x)/dx\equiv f'(x) \geq 0$. 
Since if we write the right-hand-side of the relations (\ref{gaptwo}) 
$\sim$ (\ref{gapfour}) generically as $f(x)$ we find 
$\partial^{2}{\mbox{\Large$v$}}_{\rm T}/\partial m^{2}
\Bigr|_{m^{*}}\sim xf'(x)$. 
In addition, the global minimum condition, 
${\mbox{\Large$v$}}_{\rm T}\bigr|_{m=0}
>{\mbox{\Large$v$}}_{\rm T}\bigr|_{m^{*}}$, 
must be satisfied in order to assign $m^{*}$ as the true minimum; 
since it may happen that $\bigl. {\mbox{\Large$v$}}_{\rm T} \bigr|_{m=0} 
\leq \bigl. {\mbox{\Large$v$}}_{\rm T} \bigr|_{m^{*}} $ 
under the influence of the external field.

\begin{figure}[tb]
\centering
(a)
\includegraphics*[scale=0.2]{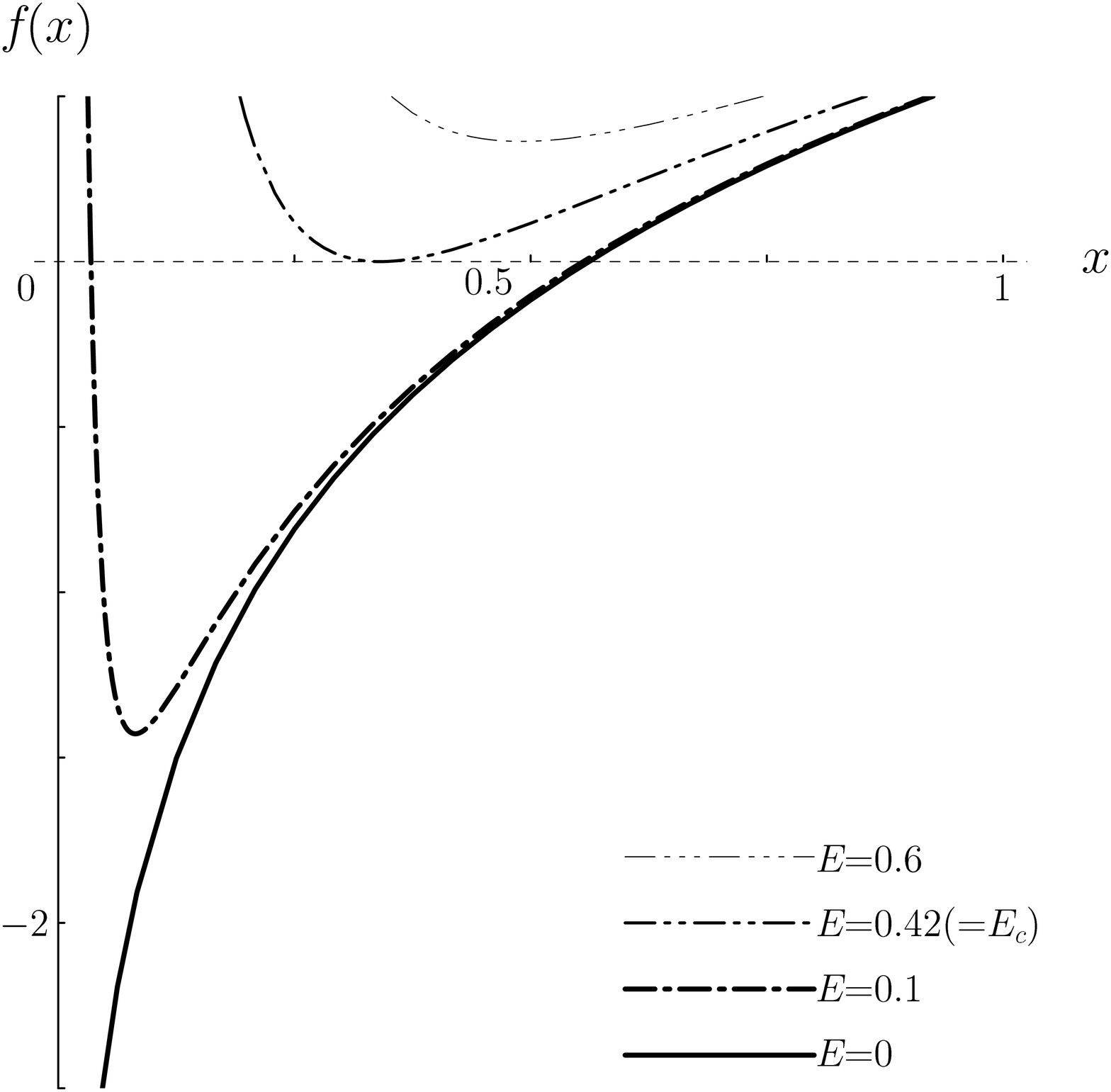}
(b)
\includegraphics*[scale=0.2]{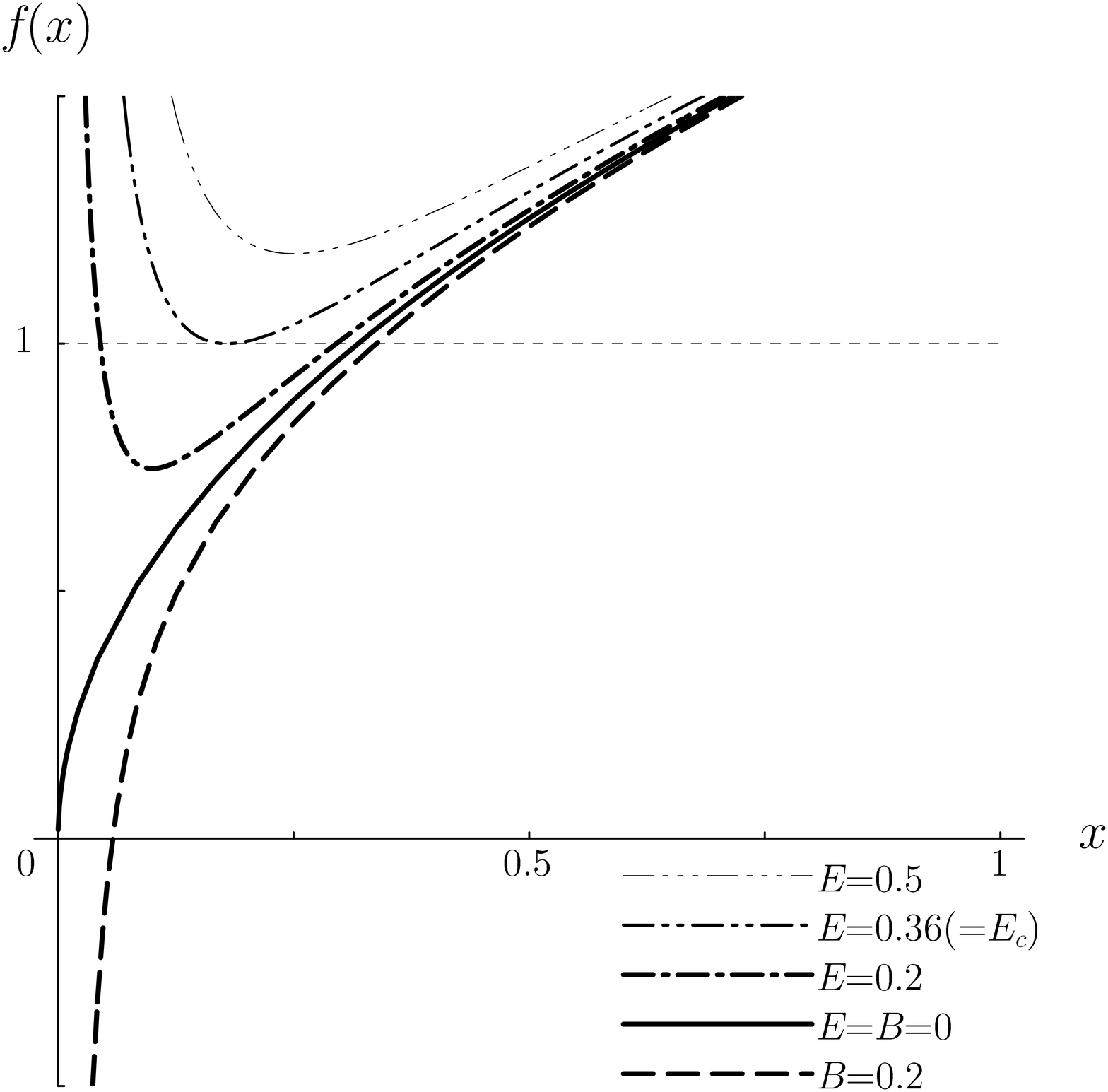}\\
\vspace{30pt}
(c)
\includegraphics*[scale=0.2]{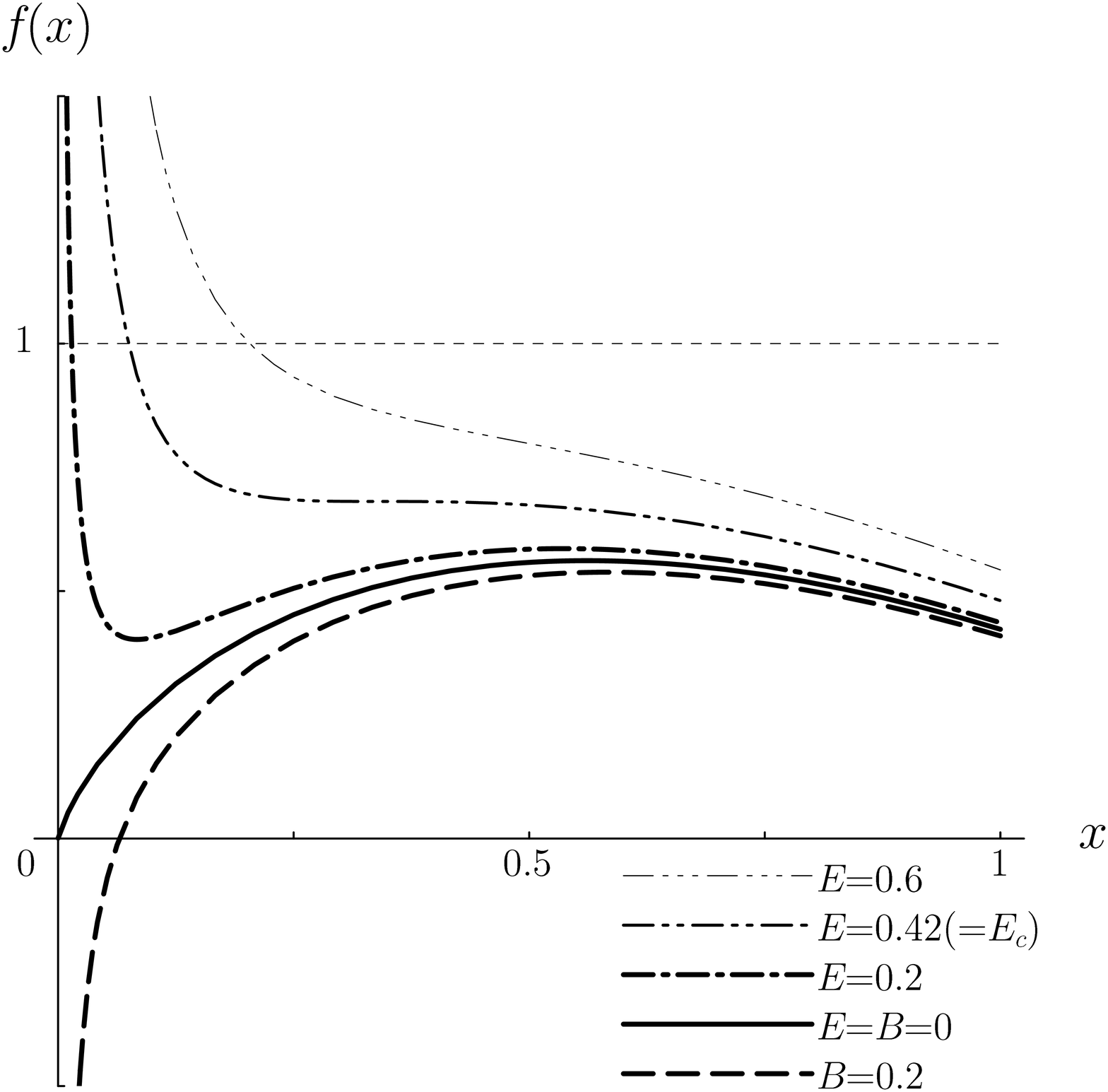}

\caption{(a), (b), and (c): the plot of the right hand sides of 
(\ref{gaptwo}), (\ref{gapthree}), and (\ref{gapfour}). 
${\mathcal F}_{D}$ is read as ${\mathcal F}_{D}=-iE$ and ${\mathcal F}_{D}=B$
. Solutions of the gap equations are found from the intersection 
of $f(x)$ and the horizontal line depicting the coupling dependence. From 
(\ref{gaptwo}) $\sim$ (\ref{gapfour}) and the stability condition, 
such intersections should be in the region: (i) under 
thin-dashed horizontal line (at $0$ in two dimensions and $1$ in 
three and four dimensions), (ii) of $f(x)$ with 
non-negative gradient. 
The critical electric fields are defined by those over which no 
chiral symmetry breaking occurs in any coupling, 
and characterized by upper curves 
attaching to the thin-dashed horizontal line in two and 
three dimensions or by the minimum curve of monotonically decreasing
in four dimensions. 
Due to the infrared cut-off, 
$m^{*2}=0$ does not correspond to $x=0$ but $x=\epsilon/\Lambda^{2}$. 
Note that the critical coupling $g_{c}$ remains finite against 
$B\ne0$ as long as $\epsilon/\Lambda^{2}$ is finite. }
\label{fx}
\end{figure}

Let us consider specific cases: a purely magnetic as well as electric 
field case. We first start with the magnetic field case, that is in 
three and four dimensions: $F 
\mapsto |\mbit{B}|$ in (\ref{gapthree}) and (\ref{gapfour}). From Figure.\ref{fx}(b),(c) 
it can be recognized that the magnetic field reduces the 
critical coupling $g_{c}$. It should be noted that
even in three dimensions {\em $g_{c}$ never reaches zero.}  
If we take $\epsilon \rightarrow 0$, while keeping $|\mbit{B}| \neq 0$, 
(implying $x \rightarrow 0$ in the figures), $g_{c}$ would tend to zero,
which, however, cannot be allowed as far as the external fields are
present.  In this way the observation by 
Gusynin et al. \cite{GMS2} is not true. They have simply ignored the 
infrared divergences to get the $g_{c}=0$ result and interpreted it by 
means of the dimensional reduction.   

Next we take the purely electric field case in two, three, and four 
dimensions: $F \rightarrow -i | 
\mbit{E}|$. (Recall that we have been in the Euclidean world.)
Again from Figure.\ref{fx}(a)$\sim$(c), it is apparent that stronger the electric field 
becomes larger the critical coupling $g_{c}$. Electric fields thus
restore the symmetry 
when they overreach 
the point $| \mbit{E}_{c}| $ (whose value is seen in the figures).  
The observation is consistent with Klevansky et al. \cite{KL} (but 
they have also ignored the infrared divergences.)

\begin{figure}[tb]
\centering
\includegraphics*[scale=0.2]{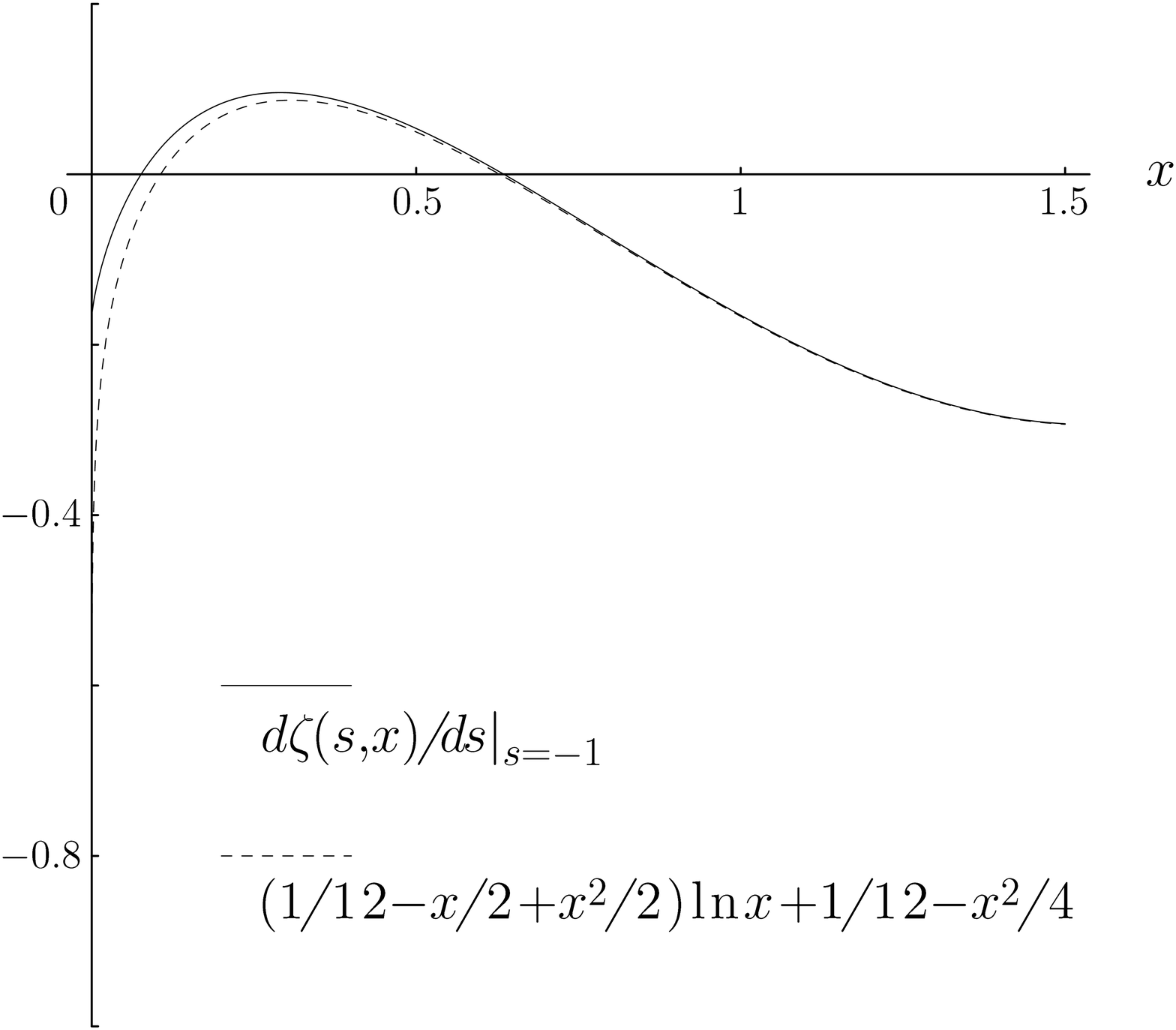}
\caption{The asymptotic expansion and the exact curve in four 
dimensions: the dashed (solid) curve is the asymptotic 
expansion ($\zeta'(-1,x)$).  Note that the matching is excellent
over the 
whole region except the origin.}
\label{zetaasymptotic}
\end{figure}

\begin{figure}[tb]
\centering
(a)\hspace{-40pt}
\includegraphics*[scale=0.2]{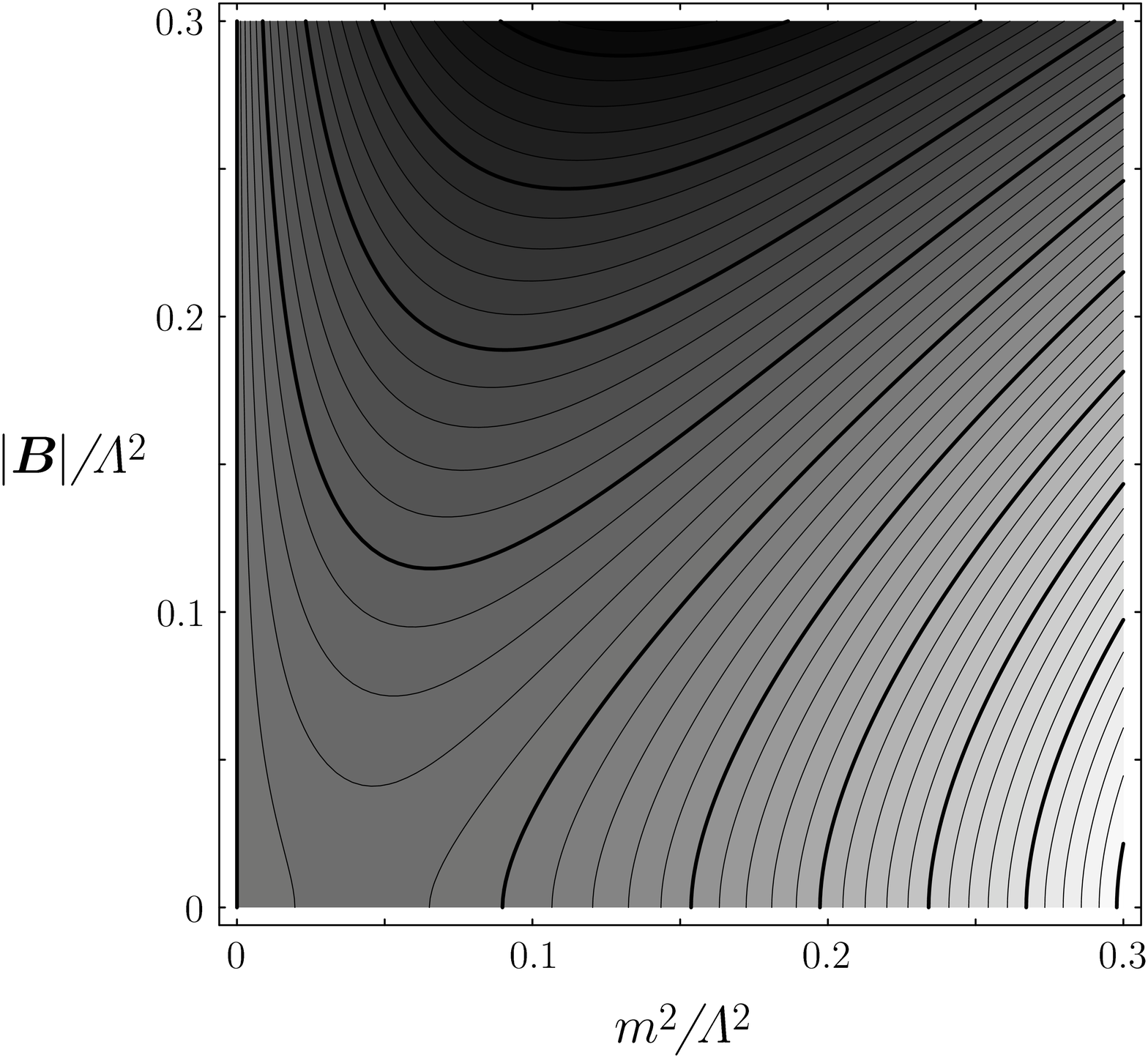}\hspace{50pt}
(b)\hspace{-40pt}
\includegraphics*[scale=0.2]{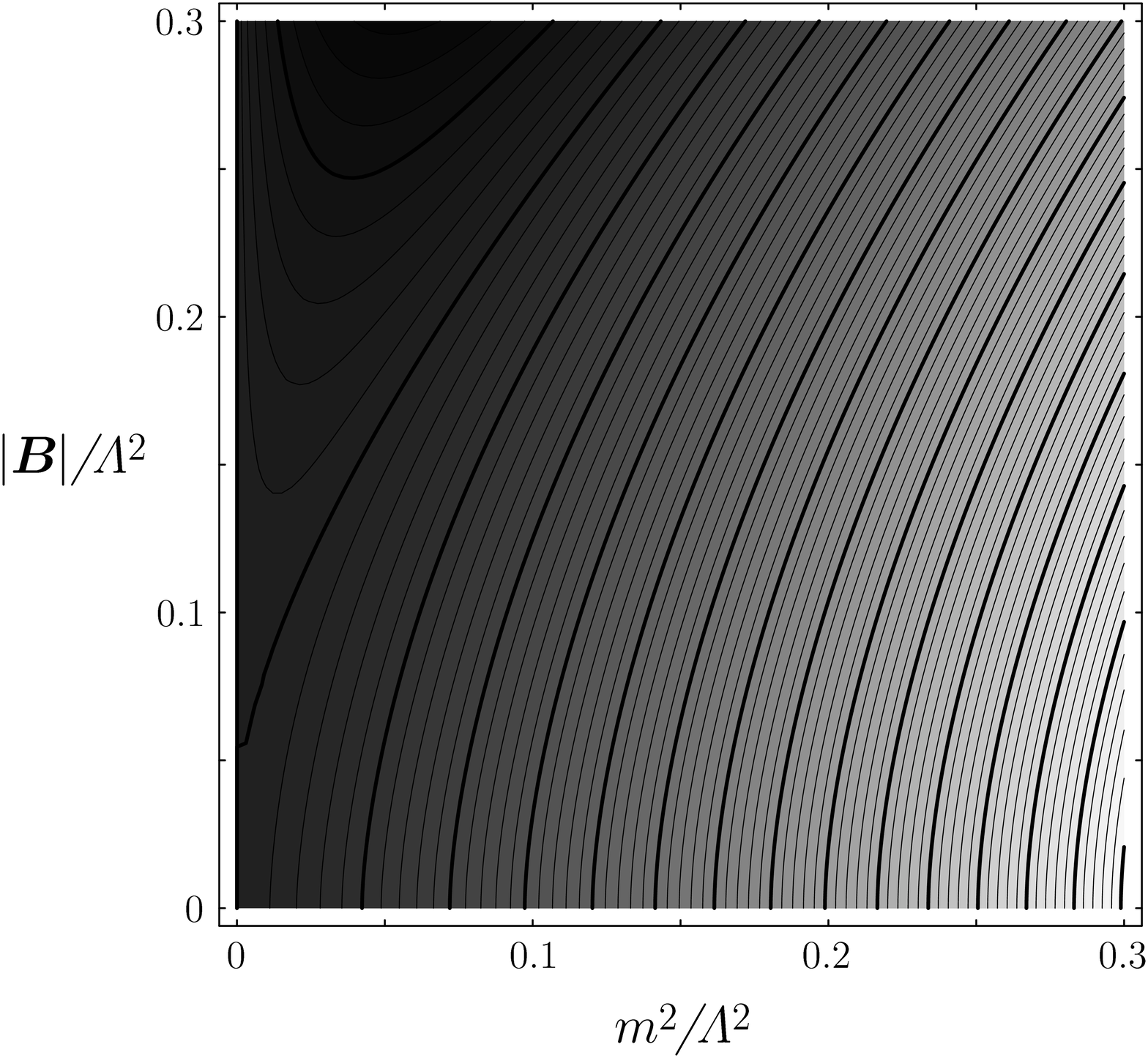}
\caption{The contour plots of the total potential with 
magnetic field (a) at $4\pi^{2}/g^{2}\Lambda^{2}=0.85$ and 
(b) at $4\pi^{2}/g^{2}\Lambda^{2}=1.15$ in four dimensions. 
The potential has been adjusted to vanish at $m^{2}/\Lambda^{2}=0$
and drawn in the unit of $\Lambda^{4}/8\pi^{2}$. 
Note that the adjustment is possible 
due to the infrared cut-off $\epsilon$ whose value is 
$\epsilon/\Lambda^{2}=0.00001$.  The thick 
contours imply the height difference of $0.01$. The minimum of the 
potential moves to the right as the magnetic field becomes stronger.}
\label{pot3d}
\end{figure}

So far we have utilized the asymptotic expansion to the potential 
(\ref{pottwo}) to (\ref{potfour}):
$a \equiv (m^{2} + \epsilon)/ 2F_{D} \sim \infty$. Then it seems that 
the discussions above are plausible only in the region where the strength
 of the external fields is tiny compared to the (generated) mass. However as 
can be seen from Figure.\ref{zetaasymptotic} (of four dimensions), the asymptotic expansion,  
(\ref{aspfour}), matches with the exact value 
up to $a \sim 0.5$ and, moreover, does not deviate from it
except around the origin. (The situation is the same for two 
and three dimensions.) Therefore the exact form of the potential in  
four dimensions depicted as Figure.\ref{pot3d} remains almost unchanged 
after employing the asymptotic expansion and we can recognize that
our minimum is indeed the global minimum. 

Accordingly we can increase external fields larger than 
the mass even under the asymptotic expansion. It should, however, 
bear in mind
that an arbitrarily large electric field cannot be 
allowed; since there emerges an 
imaginary part in the potential \cite{KL} when $a \rightarrow 0$. Intuitively 
speaking when the electric field exceeds the threshold of the 
particle, $|\mbit{E}| > m^{2}$,  a pair creation occurs, which
leads to instability of the vacuum. (The phenomena 
is closely related to the Klein paradox and is well-known \cite{GMR}.)

In four dimensions, we have assumed 
$\mbit{B} \cdot \mbit{E} =0$ so that we need not worry 
about the chiral anomaly which, however, must be
taken into account in the non-abelian case. Therefore the full 
calculation to (\ref{fullpot}) is desirable and will be seen in our 
next work \cite{IKT}. 

\vspace{3ex}  
\centerline{\bf Acknowledgment}
\noindent The authors thank to Kenzo Inoue and Koji Harada
 for valuable 
discussions.

\end{document}